\documentclass[12pt]{iopart}
\usepackage{graphicx}
\usepackage{picinpar}

\begin{document}

\title{Creating quantum discord through local generalized amplitude damping}

\author{Jianwei Xu}

\address{Key Laboratory for Radiation Physics and Technology, Institute of Nuclear Science and Technology, Sichuan University,
Chengdu 610065, China}
\ead{xxujianwei@yahoo.cn}
\begin{abstract}
We show that two qubits initially in completely classical state
can create quantum discord through a local generalized amplitude damping
channel, but high temperature will impede the creating of quantum discord.
\end{abstract}


\section{Introduction}
Quantum entanglement is one of the most striking features of quantum physics.
It remarkably reveals nonlocality and may lead to powerful applications in
quantum information and quantum computation
\cite{Nielson2000,Horodecki2009}. But entanglement is incredibly fragile
under environmental noises, and any local operations can not increase it.
These are serious frustrations for using entanglement.

Quantum discord \cite{Ollivier2001,Henderson2001} is another quantum correlation other than entanglement, and
also can be used for quantum computation \cite{Datta2008,Lanyon2008,Datta2009}. It
has been shown that quantum discord is more robust than entanglement \cite{Werlang2009}. Furthermore, recent studies
\cite{Merali2011,Ciccarello2011,Streltsov2011} indicate that local environmental noises can create quantum
discord. These are exciting results compared to the case of entanglement.

In \cite{Ciccarello2011}, F. Ciccarello and V. Giovannetti showed that a zero discord state of two qubits
\begin{eqnarray}
\tau _{ini}=\frac 12(|+\rangle \langle +|\otimes |0\rangle \langle 0|+|-\rangle \langle -|\otimes |1\rangle \langle 1|)
\end{eqnarray}
can create quantum discord under an amplitude damping channel, where
\begin{eqnarray}
|\pm \rangle =\frac 1{\sqrt{2}}(|0\rangle \pm |1\rangle ).
\end{eqnarray}
In this paper we will investigate the more general case that the initial
state is
\begin{eqnarray}
\rho _{ini}=\frac 12(|\psi _0(\lambda )\rangle \langle \psi _0(\lambda
)|\otimes |0\rangle \langle 0|+|\psi _1(\lambda )\rangle \langle \psi
_1(\lambda )|\otimes |1\rangle \langle 1|),
\end{eqnarray}
where
\noindent
\begin{eqnarray}
\fl
\ \ \ \ \  \ \ \ \  |\psi _0(\lambda )\rangle =\sqrt{\lambda }|0\rangle +\sqrt{1-\lambda }%
|1\rangle,  \ \ \
|\psi _1(\lambda )\rangle =\sqrt{1-\lambda }|0\rangle -\sqrt{\lambda }%
|1\rangle, \ \ \
\lambda \in [0,1],
\end{eqnarray}
and the noise is the generalized amplitude damping channel.
 We know generalized amplitude
damping describes temperature effects, so we are especially interested in the
temperature effect on the quantum discord. This paper is organized as
follows. In Sec. 2, we briefly recall some basics about quantum discord and
generalized amplitude damping as preparations. In Sec. 3, we investigate
the quantum discord of two qubits whose intial state as in Eq.(3) and undergoes a local
generalized amplitude damping. Sec. 4 is a brief summary.

\section{Quantum discord and generalized amplitude damping}

Suppose two quantum systems A and B are described by the Hilbert spaces $H^A$ and
$H^B$, respectively. The composite system AB is then described by the
Hilbert space $H^A\otimes $ $H^B$. For a state $\rho $ on $H^A\otimes $ $H^B$%
, the quantum discord (with respect to A) of $\rho $ is defined as
\cite{Ollivier2001}
\begin{eqnarray}
D_A(\rho )=S(\rho ^A)-S(\rho )+\inf_{\{\Pi _i\}_i}\sum_ip_iS(\rho _i^B),
\end{eqnarray}
where $\rho ^A=tr_B\rho $, $\Pi _i=|i\rangle \langle i|$, inf takes all
projective measurements $\{\Pi _i\}_i$ on system A, $p_i=tr_B\langle i|\rho
|i\rangle $, $\rho _i^B=\langle i|\rho |i\rangle /p_i$, $S(\cdot )$ is the
Von Neumann entropy. It can be proven that \cite{Ollivier2001}
\begin{eqnarray}
D_A(\rho )=0\Longleftrightarrow \rho =\sum_ip_i|i\rangle \langle i|\otimes
\rho _i^B,
\end{eqnarray}
where $\{|i\rangle \}_i$ is an orthonormal basis for $H^A$, $\{\rho _i^B\}_i$
are density operators on $H^B$, $p_i\geq 0$, $\sum_ip_i=1$.

Eq.(5) is hard to optimize, even for two qubits case till now only few states
are found possessing analytical expressions \cite{Luo2008,Ali2010}. B. Dakic, V. Vedral and C. Brukner
 proposed a geometric measure of quantum discord as \cite{Dakic2010}
\begin{eqnarray}
D_A^G(\rho )=\inf_\sigma \{tr[(\rho -\sigma )^2]:D_A(\sigma )=0\}.
\end{eqnarray}
Evidently,
\begin{eqnarray}
D_A(\rho )=0\Longleftrightarrow D_A^G(\rho )=0.
\end{eqnarray}
As an elegant result, $D_A^G(\rho )$ allows analytical expressions for all
two qubits states \cite{Dakic2010}. More specifically, for a two qubits state
\begin{eqnarray}
\rho =\frac 14(I\otimes I+\sum_{i=1}^3x_i\sigma _i\otimes
I+\sum_{j=1}^3y_jI\otimes \sigma _j+\sum_{i,j=1}^3T_{ij}\sigma _i\otimes
\sigma _j),
\end{eqnarray}
where, $\sigma _1=\sigma _x$, $\sigma _2=\sigma _y$, $\sigma _3=\sigma _z$
are Pauli matrices, $\{x_i\}_{i=1}^3$, $\{y_j\}_{j=1}^3$, $%
\{T_{ij}\}_{i,j=1}^3$ are all real numbers with
\begin{eqnarray}
x_i=tr[\rho \sigma _i\otimes I],y_j=tr[\rho I\otimes \sigma
_j],T_{ij}=tr[\rho \sigma _i\otimes \sigma _j],
\end{eqnarray}
then \cite{Dakic2010}
\begin{eqnarray}
D_A^G(\rho )=\frac 14(\sum_{i=1}^3x_i^2+\sum_{i,j=1}^3T_{ij}^2-\lambda
_{\max }).
\end{eqnarray}
Where $\lambda _{\max }$ is the largest eigenvalue of the matrix $xx^t+TT^t$, $%
x=(x_1,x_2,x_3)^t$, $T$ is the matrix $(T_{ij})$.

Generalized amplitude damping describes the effect of dissipation to an
environment at finite temperature (\cite{Nielson2000}, 8.3.5). Suppose two qubits systems A and B, given
the bases $\{|0\rangle ,|1\rangle \}$, $\{|0\rangle ,|1\rangle \}$ for $H^A$
and $H^B$, the operation elements for generalized amplitude damping are
\begin{eqnarray}
E_0=\sqrt{p}\left(
\begin{array}{ll}
1 & 0 \\
0 & \sqrt{1-\gamma }
\end{array}
\right), \ \ \ \ \ \ \   E_1=\sqrt{p}\left(
\begin{array}{ll}
0 & \sqrt{\gamma } \\
0 & 0
\end{array}
\right),    \nonumber \\
E_2=\sqrt{1-p}\left(
\begin{array}{ll}
\sqrt{1-\gamma } & 0 \\
0 & 1
\end{array}
\right), \ \
E_3=\sqrt{1-p}\left(
\begin{array}{ll}
0 & 0 \\
\sqrt{\gamma } & 0
\end{array}
\right).
\end{eqnarray}
Where $p\in [0,1]$, $\gamma \in [0,1]$. $p=p(T)$ and $\gamma =\gamma (T)$
are functions of temperature $T$. $\gamma (T)$ describes the transition
probability between $|0\rangle $ and $|1\rangle $ at temperature $T$. $p(T)$
describes temperature effects, $p(0)=1$ is the case of amplitude damping. We
can assume that (see for example \cite{Srikanth2008})  $p(T)$ is
a monotonically decreasing function when T varies from 0 to $+\infty $, and
\begin{eqnarray}
lim_{T\rightarrow 0}p(T)=1,lim_{T\rightarrow +\infty }p(T)=0.5.
\end{eqnarray}
From Eq.(12) we can see that p varies from 0 to 0.5 is symmetric to that p
varies from 1 to 0.5 in the sense exchanging states $|0\rangle$ and $|1\rangle$, so we only need to consider the case  p varies
from 1 to 0.5.

A state $\rho $ after this generalized amplitude damping will be a state
\begin{eqnarray}
E(\rho )=E_0\rho E_0^{\dagger }+E_1\rho E_1^{\dagger }+E_2\rho E_2^{\dagger
}+E_3\rho E_3^{\dagger }.
\end{eqnarray}

\section{Creating quantum discord through local generalized amplitude damping}

Now suppose two qubits initially is in the state as in Eq.(3),
through the generalized amplitude damping channel in Eq.(12), the output state $\rho=E(\rho_{ini})$
can be calculated directly by Eqs.(3,12,14),  will be
\begin{eqnarray}
\rho =\frac 14(I\otimes I+x_3\sigma _3\otimes I+T_{13}\sigma _1\otimes
\sigma _3+T_{33}\sigma _3\otimes \sigma _3),
\end{eqnarray}
where
\begin{eqnarray}
x_3=(2p-1)\gamma , \ \  T_{13}=2\sqrt{\lambda (1-\lambda )(1-\gamma )}%
, \ \  T_{33}=(1-\gamma )(2\lambda -1).
\end{eqnarray}
We first ask, for what initial states and what generalized amplitude damping
channel, the quantum discord of the output states, $D_A(\rho )$, are vanishing? This
is the Proposition 1 below.

\emph{Proposition 1.} The quantum discord $D_A(\rho )$ of the state $\rho $ in
Eq.(15) is zero if one of the conditions below is satisfied

(i).  $\gamma =0;$

(ii).  $\gamma =1;$

(iii).  $\lambda =0;$

(iv).  $\lambda =1;$

(v).  $p=0.5.$

\emph{Proof.} From Eq.(8), we only need to prove $D_A^G(\rho )=0.$ Using Eq.(11) to
Eq.(15), after some calculations, we get
\begin{eqnarray}
\fl  \ \ \ \ \ \ \ \
D_A^G(\rho )=\frac 14\{(x_3^2+T_{13}^2+T_{33}^2)-\sqrt{%
[(x_3+T_{13})^2+T_{33}^2][(x_3-T_{13})^2+T_{33}^2]}\}.
\end{eqnarray}
From Eq.(16), it is easy to see that

(vi). if $\gamma =0$, then $x_3=0;$

(vii). if $\gamma =1$, then $T_{13}=T_{33}=0;$

(viii). if $\lambda =0$, then $T_{13}=0;$

(ix). if $\lambda =1$, then $T_{13}=0;$

(x). if $p=0.5$, then $x_3=0$.

Each of the cases above can result in $D_A^G(\rho )=0$. We then complete
this proof.

We next investigate the quantum discord $D_A(\rho )$ for the state $\rho $
in Eq.(15) for more general cases, that is $\gamma \in (0,1)$, $\lambda \in
(0,1)$, $p\in (0.5,1]$.

Any orthonormal basis of $H^A$ can be expressed as an unitary matrix $U$
multiplied by the orthonormal basis $\{|0\rangle ,|1\rangle \}$. Since $U$
can be written as
\begin{eqnarray}
U=\left(
\begin{array}{ll}
t+ic & -b+ia \\
b+ia & t-ic
\end{array}
\right),
\end{eqnarray}
where $t,a,b,c$ are all real numbers and satisfy
\begin{eqnarray}
t^2+a^2+b^2+c^2=1,
\end{eqnarray}
then any orthonormal basis of $H^A$ can be expressed as
\begin{eqnarray}
|\varphi _0\rangle =(t+ic)|0\rangle +(b+ia)|1\rangle ,  \nonumber \\
|\varphi _1\rangle =(-b+ia)|0\rangle +(t-ic)|1\rangle .
\end{eqnarray}
Notice that
\begin{eqnarray}
I|0\rangle =|0\rangle , \ \ \  \sigma _x|0\rangle =|1\rangle ,  \ \ \  \sigma _y|0\rangle
=i|1\rangle ,  \ \ \ \sigma _z|0\rangle =|0\rangle , \nonumber \\
I|1\rangle =|1\rangle , \ \ \ \sigma _x|1\rangle =|0\rangle , \ \ \ \sigma _y|1\rangle
=-i|0\rangle , \ \sigma _z|1\rangle =-|1\rangle .
\end{eqnarray}
Denote $p_0\rho _0=\langle \varphi _0|\rho |\varphi _0\rangle $ and $p_1\rho _1=\langle \varphi _1|\rho |\varphi _1\rangle$, then from Eqs.(15,20,21), we have
\begin{eqnarray}
\fl
p_0\rho _0=\frac I4[1+(t^2+c^2-a^2-b^2)x_3]+\frac{\sigma _z}
4[2(tb+ac)T_{13}+(t^2+c^2-a^2-b^2)T_{33}],    \\
\fl
p_0=tr\langle \varphi _0|\rho |\varphi _0\rangle =\frac
12(1+x_3)-(a^2+b^2)x_3,   \\
\fl
p_1\rho _1=\frac I4[1+(a^2+b^2-t^2-c^2)x_3]+\frac{\sigma _z}
4[-2(tb+ac)T_{13}+(a^2+b^2-t^2-c^2)T_{33}],    \\
\fl
p_1=tr\langle \varphi _1|\rho |\varphi _1\rangle =\frac
12(1-x_3)+(a^2+b^2)x_3.
\end{eqnarray}
Let
\begin{eqnarray}
\fl   \ \ \ \ \ \ \ \
a=\sqrt{r}cos\theta _1, \ \  b=\sqrt{r}\sin \theta _1, \ \ c=\sqrt{1-r}\sin \theta
_2, \ \ t=\sqrt{1-r}\cos \theta _2,
\end{eqnarray}
then
\begin{eqnarray}
a^2+b^2=r, \ \ \ tb+ac=\sqrt{r(1-r)}\sin \theta _3,
\end{eqnarray}
where
\begin{eqnarray}
r\in [0,1],  \ \ \   \theta _1\in [0,2\pi ),   \ \ \  \theta _2\in [0,2\pi ),  \ \ \  \theta
_3=\theta _1+\theta _2.
\end{eqnarray}
Eqs.(22-25) then read
\begin{eqnarray}
\fl \ \ \ \ \ \ \ \
p_0\rho _0=\frac I4[1+(1-2r)x_3]+\frac{\sigma _z}4[2\sqrt{r(1-r)}\sin
\theta _3T_{13}+(1-2r)T_{33}],  \\
\fl \ \ \ \ \ \ \ \
p_0=\frac 12[1+(1-2r)x_3],   \\
\fl \ \ \ \ \ \ \ \
p_1\rho _1=\frac I4[1-(1-2r)x_3]+\frac{\sigma _z}4[-2\sqrt{r(1-r)}\sin
\theta _3T_{13}-(1-2r)T_{33}],  \\
\fl \ \ \ \ \ \ \ \
p_1=\frac 12[1-(1-2r)x_3].
\end{eqnarray}
Let
\begin{eqnarray}
\alpha =1-2r,   \ \ \ \   \beta =2\sqrt{r(1-r)}\sin \theta _3,
\end{eqnarray}
then Eqs.(29-32) become
\begin{eqnarray}
p_0\rho _0=\frac I4[1+\alpha x_3]+\frac{\sigma _z}4[\beta T_{13}+\alpha
T_{33}],   \\
p_0=\frac 12[1+\alpha x_3],   \\
p_1\rho _1=\frac I4[1-\alpha x_3]+\frac{\sigma _z}4[-\beta T_{13}-\alpha
T_{33}],   \\
p_1=\frac 12[1-\alpha x_3],
\end{eqnarray}
with
\begin{eqnarray}
\alpha ^2+\beta ^2\leq 1.
\end{eqnarray}
From Eqs.(34-37), it is easy to find that $\rho _0$ has two eigenvalues
\begin{eqnarray}
\frac{1+\alpha (x_3+T_{33})+\beta T_{13}}{2(1+\alpha x_3)}, \ \ \ 1-\frac{1+\alpha
(x_3+T_{33})+\beta T_{13}}{2(1+\alpha x_3)},
\end{eqnarray}
and $\rho _1$ has two eigenvalues
\begin{eqnarray}
\frac{1-\alpha (x_3+T_{33})-\beta T_{13}}{2(1-\alpha x_3)}, \ \ \ 1-\frac{1-\alpha
(x_3+T_{33})-\beta T_{13}}{2(1-\alpha x_3)}.
\end{eqnarray}
Thus
\begin{eqnarray}
\fl  \ \ \
p_0S(\rho _0)+p_1S(\rho _1)=F(\alpha ,\beta )   \nonumber \\
\fl
=\frac 12(1+\alpha x_3)h(\frac{1+\alpha (x_3+T_{33})+\beta T_{13}}{2(1+\alpha x_3)})+\frac 12(1-\alpha x_3)h(\frac{1-\alpha (x_3+T_{33})-\beta T_{13}}{2(1-\alpha x_3)}),
\end{eqnarray}
where $h(\cdot )$ is the binary entropy
\begin{eqnarray}
h(x)=-xlog_2x-(1-x)\log _2(1-x),
\end{eqnarray}
$h(x)$ is defined on $x\in [0,1],$ $h(x)$ is a concave function, and $%
h(x)=h(1-x).$

We say $F(\alpha ,\beta )$ is a concave function in $(\alpha,\beta)$, this can be seen from the concavity of $h(x)$  and the
facts below:

(xi). If $f(x)$ and $g(x)$ are convex functions, then so is $f(x)+g(x)$
(\cite{Boyd2004}, 3.2.1);

(xii). Convexity is invariant under affine maps: that is, if f(x) is convex
with $x\in R^n$, then so is $g(y)=f(Ay+x_0)$ with $y\in R^m$, $x_0\in R^n$, $%
A$  an $n\times m$ real matrix (\cite{Boyd2004}, 3.2.2);

(xiii). If f(x) is convex, then so is $g(x,t)=tf(x/t)$ with $t>0$
(\cite{Boyd2004}, 3.2.6).

It follows that the minimum of $F(\alpha ,\beta )$ in Eq.(41) over the domain $\alpha
^2+\beta ^2\leq 1$ can be achieved on the unit circle $\alpha ^2+\beta ^2=1.$
Further, from Eq.(41), it is evident that
\begin{eqnarray}
F(\alpha ,\beta )=F(-\alpha ,-\beta ),
\end{eqnarray}
hence,
\begin{eqnarray}
\min [p_0S(\rho _0)+p_1S(\rho _1):all \ \{\Pi_{i}\}_{i=1}^{2}]=\min \{F(\theta ):0\leq \theta \leq \pi \},
\end{eqnarray}
where
\begin{eqnarray}
F(\theta )=\frac 12(1+x_3\cos \theta )h(\frac{1+(x_3+T_{33})\cos \theta
+T_{13}\sin \theta }{2(1+x_3\cos \theta )})   \nonumber \\   \ \ \ \ \ \ \ \
+\frac 12(1-x_3\cos \theta )h(\frac{1-(x_3+T_{33})\cos \theta -T_{13}\sin
\theta }{2(1-x_3\cos \theta )}).
\end{eqnarray}
Finally, we get the expression of $D(\rho )$ as
\begin{eqnarray}
D(\rho )=S(\rho ^A)-S(\rho )+\min \{F(\theta ):0\leq \theta \leq \pi \},
\end{eqnarray}
where $S(\rho ^A)$ and $S(\rho )$ can be calculated directly by Eq.(15) as
\begin{eqnarray}
\fl  \ \ \ \ \ \
S(\rho ^A)=h(\frac{1+x_3}2),  \\
\fl  \ \ \ \ \ \
S(\rho )=1-\frac 12h(\frac{1+\sqrt{(x_3+T_{33})^2+T_{13}^2}}2)-\frac 12h(%
\frac{1+\sqrt{(x_3-T_{33})^2+T_{13}^2}}2).
\end{eqnarray}
From these expressions, we have Proposition 2 below.

\emph{Proposition 2.} Quantum discord $D_A(\rho )$ in Eq.(46) for the state $\rho $ in Eq.(15) has the same
value for $(p,\lambda ,\gamma )$ and $(p,1-\lambda ,\gamma )$, that is, $%
D_A(\rho )$ is symmetric about $\lambda =0.5$.

\emph{Proof.} From Eq.(16), if $(p,\lambda ,\gamma )$ generates $(x_3,T_{13},T_{33})$%
, then $(p,1-\lambda ,\gamma )$ generates $(x_3,T_{13},-T_{33})$. $%
(x_3,T_{13},T_{33})$ and $(x_3,T_{13},-T_{33})$ generate the same value for $%
S(\rho ^A)-S(\rho )$, see Eqs.(47-48). Also, in $F(\theta )$ of Eq.(45), by use of $%
h(x)=h(1-x)$, we can get
\begin{eqnarray}
F(\theta ,x_3,T_{13},T_{33})=F(\pi -\theta ,x_3,T_{13},-T_{33}).
\end{eqnarray}
Thus, we can readily attain Proposition 2 and end this proof.

\includegraphics[width=7cm,trim=0 0 0 0]{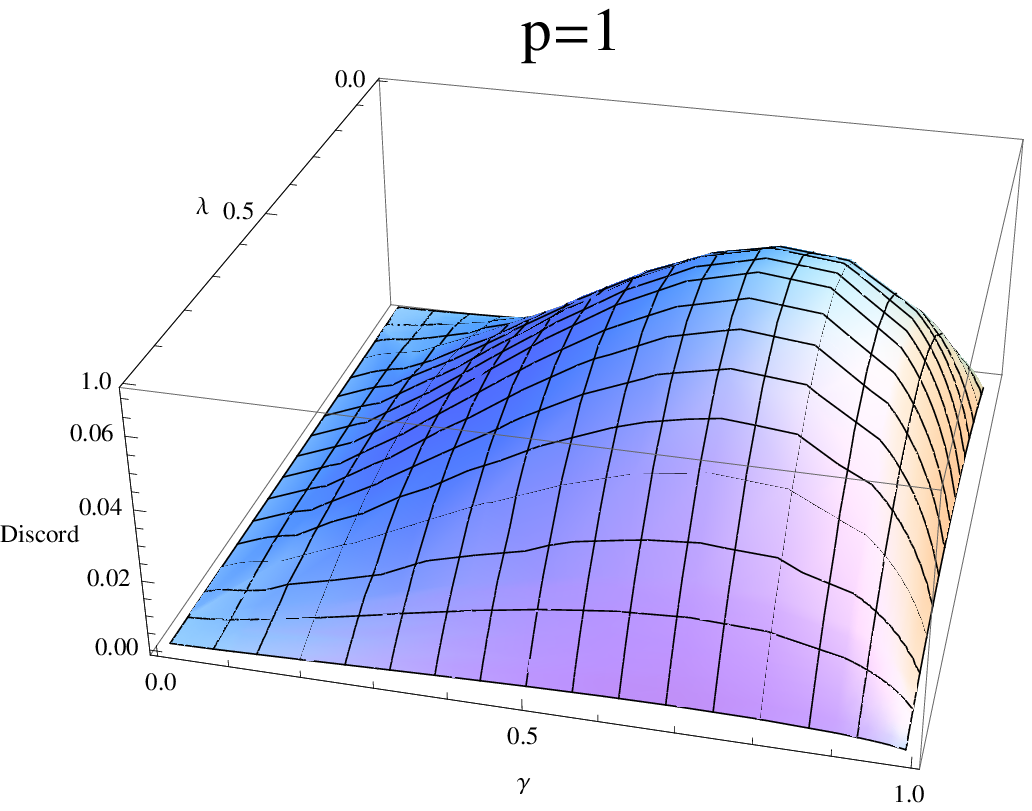}
\includegraphics[width=7cm,trim=0 0 0 0]{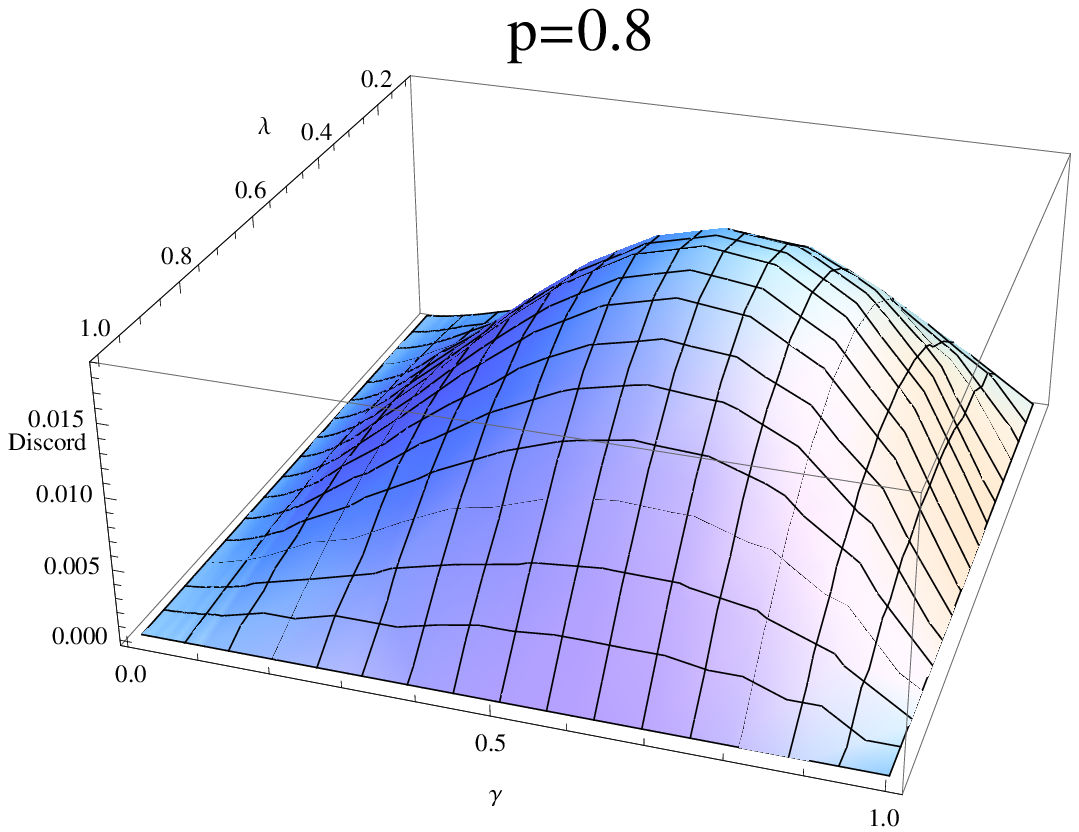}
\begin{figure}[!ht]
\includegraphics[width=7cm,trim=0 0 0 0]{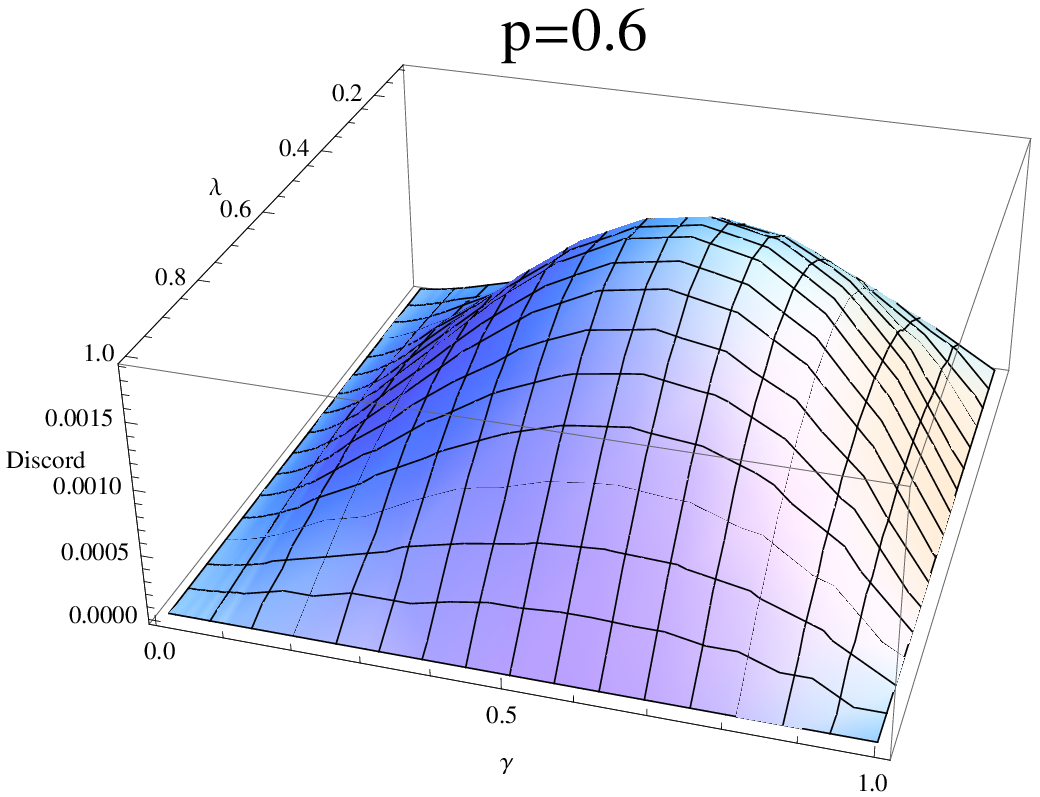}
\includegraphics[width=7cm,trim=0 0 0 0]{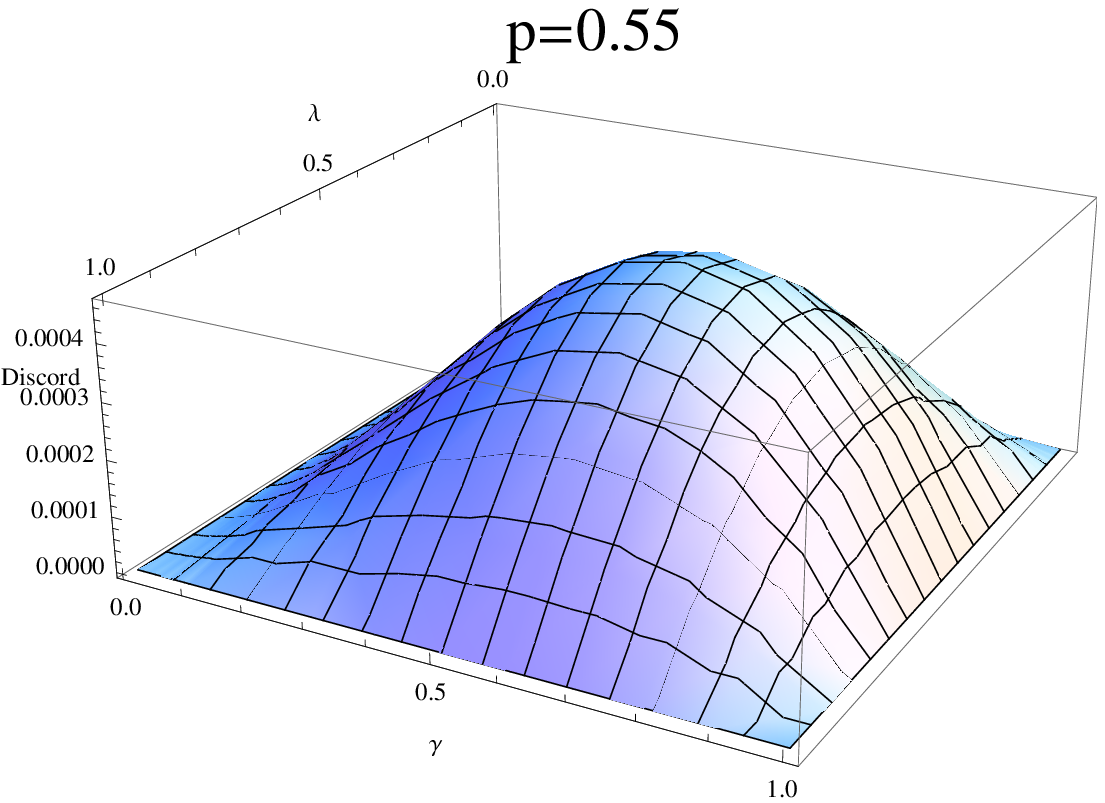}
\caption{Discord versus $\lambda$ and $\gamma$ when $p=1, p=0.8, p=0.6, p=0.55$.}
\end{figure}

\begin{window}[0,r,{\includegraphics[width=7cm,trim=0 0 0 0]{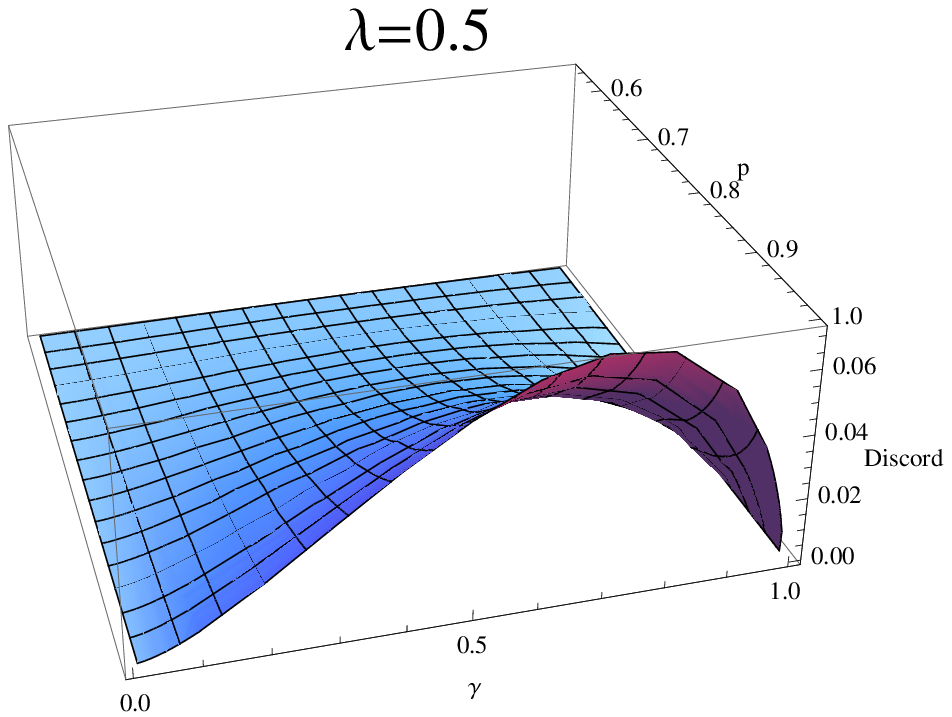}},{\textbf{Figure 2.} Discord versus $p$ and $\gamma$ when $\lambda=0.5$.}]
Fig.1 depicts $D_A(\rho )$ as a function of $(\lambda ,\gamma )$ at $p=1$, $p=0.8$, $p=0.6$, $p=0.55$, respectively. From Fig.1 we see
that, when $\lambda $ is close to $\lambda =0.5$ and $p$ close to 1, $%
D_A(\rho )$ is relatively large. That is to say, more superposition of the
states $|0\rangle $ and $|1\rangle $ in Eq.(3) can boost the creating of
quantum discord, while high temperature, conversely, can impede the creating. Fig.2
shows $D_A(\rho )$ as a function of $(p,\gamma )$ at $\lambda =0.5$.
\end{window}

\section{Summary}

We investigated the quantum discord of two qubits which initially in completely classical state then
experienced a local generalized amplitude damping channel. We showed that a
completely classical state can create quantum discord through a local
generalized amplitude damping, while high temperature will impede the
creating of quantum discord.

\section*{Acknowledgements}

This work was supported by National Natural Science Foundation of China
(Grant Nos. 10775101). The author thanks Qing Hou for helpful discussions.

\end{document}